\documentclass[aps,pra,twocolumn,showpacs]{revtex4}
\usepackage{amsbsy,latexsym,amsmath}
\usepackage{amsfonts}
\usepackage{amssymb}
\usepackage[mathscr]{eucal}
\usepackage{graphics}

\newcommand{\la}{\lambda}

\newcommand{\tr}{{\rm tr}}

\newcommand{\Blmda}{{\boldsymbol \lambda}}

\newcommand{\xt}{{\boldsymbol X}}

\newcommand{\ket}[1]{\left|{#1}\right\rangle}
\newcommand{\bra}[1]{\left\langle{#1}\right|}

\newcommand{\ketbra}[2]{\left|{#1}\rangle\!\langle{#2}\right|}
\newcommand{\mean}[1]{\langle{#1}\rangle}
\newcommand{\be}{\begin{equation}}
\newcommand{\ee}{\end{equation}}

\begin{document}
\title{Estimation of quantum finite mixtures}
\author{J.~I.~de~Vicente, J.~Calsamiglia, R.~Mu{\~n}oz-Tapia and E.~Bagan}
\affiliation{Grup de F{\'\i}sica Te{\`o}rica, Facultat de
Ci{\`e}ncies, Edifici Cn, Universitat Aut{\`o}noma de Barcelona, 08193
Bellaterra (Barcelona) Spain}
%\date{\today}

\begin{abstract}
We consider the problem of determining the weights of a quantum ensemble.  That is to say, given a quantum system that is in a set of possible {\em known} states according to an {\em unknown} probability law, we give strategies to estimate the individual probabilities, weights, or mixing proportions.
Such~strategies can be used to estimate the frequencies at which different independent signals are emitted by a source. They can also be used to estimate the weights
of particular terms in a canonical decomposition of a quantum channel.
The quality of these strategies is quantified by a covariance-type error matrix. According with this cost function, we give optimal strategies in
both the single-shot and multiple-copy scenarios. The latter is also analyzed in the asymptotic limit of large number of copies.
We give closed expressions of the error matrix for two-component quantum mixtures of qubit systems. The Fisher information plays an unusual role in the problem at hand, providing exact expressions of the minimum covariance matrix for any number of copies.

\end{abstract}

\maketitle

\section{Introduction}

Suppose we are given a quantum system which is known to be in one of several states with some unknown probability, such as a photon that travels through a communication channel and codifies some message.
These states can be non-orthogonal due to, e.g., errors occurring during the transmission, but can also be made to overlap intentionally, e.g., to avoid possible eavesdropper attacks in quantum key distribution.
Given this set of possible fixed states, we wish to find
an estimate of the probabilities that best describe the state we have been provided with.
More succinctly,  assuming that a state~$\rho_{\Blmda}$ is a convex combination of a given set of states $\{\rho_{r}\}$,
\be
\rho_{\Blmda}=\sum_{r=1}^M \lambda_{r} \rho_{r} ,
\label{eq:rhol}
\ee
we wish to best estimate the value of the weights $\{\lambda_r\}$, which we arrange in the column vector $\Blmda$ and characterize the
quantum ensemble $\{(\lambda_r, \rho_r)\}$, by performing suitable measurements on the system.

The analogous classical problem appears in the field of statistical modeling under the name of estimation of \emph{finite mixtures} \cite{mclachlan_finite_2000}. The formal study of finite mixtures was initiated by Pearson in 1894 \cite{pearson_1894}. He was conducting a biometric investigation on data collected from crabs, and found that the distribution of the size of their forehead (relative to the size of the body) presented an unexpected skewness, which could not be modeled with a symmetric normal distribution. Pearson showed that the data was very well fitted by a mixture of two normal distributions. The presence of two components was taken by Pearson as evidence that there were two different species of crabs. In this way finite mixture models can be used to expose any grouping in underlying data (clustering of data).
With the prior knowledge on the individual component densities, which can be inferred or estimated by other means, finite mixture estimation enables one to estimate the weights, or proportions, of the different populations from the gathered coarse-grained data.
A (classical) finite mixture, $p_{\Blmda}(i)= \sum\lambda_{r}p_{r}(i)$ (in the obvious notation), can thus always be interpreted as describing situations where the information on the grouping is lost, or in other words, as marginals of a joint distribution $p(i,r)$, such that $p_{\Blmda}(i)=\sum_{r} p(i,r)$, i.e.,
$p_{r}(i)$ can be viewed as the conditioned probability~$p_{r}(i)=p(i|r)$.

In this paper we approach the problem of estimating {\em quantum finite mixtures}. More precisely, we give optimal strategies to estimate the vector of weights $\Blmda$ under the assumptions given above (known set $\{\rho_r\}$ of possible states). We address also the situation in which we are provided with $N$ identical and independent copies of the state $\rho_\Blmda$, to which we will refer as {\em average state}. In this multiple-copy scenario, we further assume that generalized collective measurements can be performed on~$\rho_\Blmda^{\otimes N}$. For large $N$, we also give (local) strategies based on projective measurements on individual copies that have the same performance as the optimal collective strategies.

Quantum ensembles are necessary to describe situations in which complete prior  information is lacking.  In the context of quantum communication, for instance, one estimates the frequency of different (known) states coming out of a source, i.e., one gathers information from the average state in connection to its particular preparation procedure. It is well known that in general there is no unique quantum ensemble consistent with a given mixed state~\cite{hughston}. Therefore there will be instances in quantum finite mixture estimation, called unidentifiable,
where the average state $\rho_{\Blmda}$  does not fully determine the value of the weights $\Blmda$, which therefore cannot be estimated with unlimited precision even when an arbitrary number of copies of $\rho_{\Blmda}$  is provided. This problem is related to that of discrimination of quantum ensembles \cite{oreshkov}, where it is necessary to consider as inequivalent the different ensembles that are consistent with a given mixed state. We also note that, as in the classical case,  a quantum finite mixture can be interpreted as the marginal density matrix of an extended system-ancilla state when a particular measurement is done on the ancilla. A quantum ensemble also describes the output of a stochastic quantum channel (or generalized measurement) for a fixed input state. In particular, if the input state is taken to be one part of a bipartite maximally entangled state, the stochastic channel is fully characterized by the output state, and it can be interpreted as a quantum finite mixture. Therefore, the results that we present here can be applied to the estimation of the weights of the individual (or  of a sub-set of)  Kraus operators in a particular operator sum representation of a channel. For example, we can easily give bounds on the precision of estimating the weight of
bit flip, phase flip, and combined bit-phase flip errors, or also the total weight of 2-qubit Pauli errors versus single qubit Pauli errors.

Quantum finite mixture estimation is a novel ground for quantum estimation theory~\cite{Hel, Hol, MasahitoBook, paris_quantum_2004}, which is one of the basic tools in the field of quantum information and  has been continuously developing since the late 70's. Many problems have been addressed, ranging from the estimation of a single parameter ---as, e.g., a  phase~\cite{paris_quantum_2004}, or the losses of a quantum channel~\cite{hotta-2005}--- to full tomography.  Quantum estimation theory finds also many applications in  quantum metrology~\cite{giovannetti_quantum_2006} ---such as improvement of frequency standards~\cite{huelga_improvement_1997}, gravitational-wave detection~\cite{caves_quantum-mechanical_1981,goda_quantum-enhanced_2008},
and clock synchronization~\cite{jozsa_quantum_2000, de_burgh_quantum_2005}--- and it is often a key ingredient  in other quantum computation~\cite{cleve_quantum_1998} and communication topics, e.g., quantum benchmarks for teleportation experiments~\cite{braunstein_criteria_2000}.
The recent problem studied by Konrad {\em et al.}~\cite{Kon} can be viewed as a  quantum finite mixture estimation in a simplified context. In the present paper we address the issue in full generality. This, in passing, will enable us to  answer most of the questions posed there.

The paper is organized as follows. In Section~\ref{sec2} we introduce the general framework and give the main results for both, the single- and multiple-copy scenarios.  The asymptotic limit of large number of copies is addressed in Section~\ref{sec3}. The two sections conclude with a discussion on unidentifiability of mixtures and its consequences. Additionally, in each of these sections, we provide examples to illustrate the use of the techniques that we introduce. Section~\ref{sec4} is devoted to two-component mixtures, where closed expressions can be given for rather general situations.  The conclusions are in Section~\ref{sec5} and several technical details can be found in the appendixes, which also include an example of a two-step adaptive local strategy that is optimal.

\section{Estimation of weights in finite mixtures}\label{sec2}

\subsection{General framework}\label{ebc28.07.9S-1}

As already mentioned in the introduction, a quantum finite mixture is defined to be the convex combination  in Eq.~(\ref{eq:rhol}), where $\Blmda$ belongs to the unit ($M-1$)-simplex (i.e., the set~$\{\Blmda: \lambda_r\ge0, \sum_{r=1}^M\lambda_r=1\}$). By quantum finite mixture estimation we mean the following:
assume we have been provided with a copy of the average state~$\rho_\Blmda$ (or with several identical and independent copies of it; i.e., with $\rho_\Blmda^{\otimes N}$), of which we know nothing about the actual value of~$\Blmda$ but that it has been drawn from a (prior) probability distribution $\pi(\Blmda)$.
Assume also that we are allowed to perform generalized measurements on the copy (or copies) of~$\rho_\Blmda$.
Our task is to determine~$\Blmda$ (or, maybe, some linear combinations of its components~$\lambda_r$; namely, $a=\boldsymbol a^t\Blmda$, where $\boldsymbol a$ is some vector of constants~$a_r$). This has necessarily to be based on the output(s) of our measurement(s) on $\rho_\Blmda$ ($\rho_\Blmda^{\otimes N}$). Due to the inherent nature of quantum measurements, the determination of $\Blmda$ cannot be perfect and we can only hope to obtain an estimate within some accuracy. Our goal is to obtain the best estimate.

To give a precise meaning to the term `best estimate' we take a Bayesian approach and introduce as cost function the covariance-type error matrix
\be
\boldsymbol\Delta=\langle(\Blmda-\Blmda_\chi)(\Blmda-\Blmda_\chi)^t\rangle,
\ee
where $\Blmda_\chi$ is our estimate of $\Blmda$ based on the outcome $\chi$ of our measurement and $\langle \;\cdot\;\rangle$ stands for averaging over~$\Blmda$ and $\chi$. More precisely, the averaging is performed over the joint probability distribution $p(\chi, \Blmda)=p(\chi | \Blmda)\pi(\Blmda)$, where $p(\chi | \Blmda)$ is the probability of obtaining the outcome~$\chi$ conditioned to the actual value of~$\Blmda$. In Quantum Mechanics this conditional probability is given by Born's rule:  $p(\chi | \Blmda)=\tr\,  E_{\chi}\rho_\Blmda$, where $\{E_{\chi}\}$ is the positive operator-valued measure (POVM) that defines our generalized  quantum measurement.
The trace of the error matrix~$\boldsymbol \Delta$ gives the total mean square error (MSE), $E=\tr\boldsymbol \Delta$, while the expectation value $E_{a}=\boldsymbol a^t\boldsymbol\Delta \boldsymbol a$ gives the mean square error in the estimation of ~$a$.

In order to analyse one-copy and multiple-copy estimation in a unified framework
we have found it convenient to define quantum finite mixtures, Eq.~(\ref{eq:rhol}), in a slightly more general form, allowing for non-linear mixtures of the type
\be
\rho_{\Blmda}=\sum_{\alpha} c_{\alpha}(\Blmda)\rho_{\alpha},
\label{eq:rhog}
\ee
where the coefficient functions satisfy $\sum_{\alpha} c_{\alpha}(\Blmda)=1$ for all $\Blmda$ [but not necessarily $c_{\alpha}(\Blmda)\ge0$], and the range of values for $\alpha$ may not coincide with that for $r$ in~(\ref{eq:rhol}). As for linear finite mixtures, our goal still is to best estimate~$\Blmda$ (we assume that the functional dependence of the coefficient functions $c_\alpha$ on~$\Blmda$ is known).

The error matrix $\boldsymbol \Delta$ can be written as
\be
\boldsymbol\Delta=\sum_\chi p(\chi)\langle(\Blmda-\Blmda_\chi)(\Blmda-\Blmda_\chi)^t\rangle_\chi,
\label{eq:DeltarsG}
\ee
where $p(\chi)$ is the marginal of $p(\chi, \Blmda)$ and $\mean{\;\cdot\;}_{\chi}$ indicates averaging over the conditional probability $p(\Blmda|\chi)=p(\chi, \Blmda)/p(\chi)$ (Bayes rule). More explicitly,  $p(\chi)=\int d\Blmda\, p(\chi, \Blmda)$, where we use the shorthand notation $d\Blmda=\delta\left(\sum_r\lambda_r-1\right)\prod_r d\lambda_r$. Note that the Dirac $\delta$-function, along with $\lambda_r\ge0$, guarantees that $\Blmda$ is a point in the unit $(M-1)$-simplex (hereafter, simplex for brevity).
Eq.~(\ref{eq:DeltarsG}) can be cast as
\be
\boldsymbol\Delta=\sum_{\chi} p(\chi)\left\{ \left\langle \left(\Blmda\!-\!\mean{\Blmda}_{\chi}\right)\left(\Blmda\!-\!\mean{\Blmda}_{\chi}\right)^t\right\rangle_{\chi}\!\!+ \boldsymbol\delta_\chi \boldsymbol\delta_\chi^t\right\},
\label{ebc16.07.09-1}
\ee
with $\boldsymbol\delta_\chi=\mean{\Blmda}_{\chi}-\Blmda_\chi$. Note that all dependence on our particular choice of the estimator $\Blmda_\chi$ is contained in $\boldsymbol\delta_\chi$. Since the matrix $\boldsymbol\delta_\chi \boldsymbol\delta_\chi^t$ is manifestly positive semi-definite, the estimator that minimizes our cost function~$\boldsymbol\Delta$~is
\be
\Blmda_\chi =\mean{\Blmda}_{\chi}= \int d\Blmda\,p(\Blmda|\chi) \Blmda
={\int d\Blmda\,\pi(\Blmda)\, \Blmda\,\tr E_\chi\rho_\Blmda \over
\int d\Blmda\, \pi(\Blmda)\tr E_\chi\rho_\Blmda}
 .  \label{eq:optguess}
\ee
(Note that the components of $\Blmda_\chi$ are non-negative and add up to one; i.e., $\Blmda_\chi$ is a probability vector)
Hereafter, we will only consider this optimal estimator, which gives the smallest error matrix.  We will denote this matrix  by the same symbol $\boldsymbol\Delta$ to simplify the notation. Hence, we may write
\be
\boldsymbol\Delta = \sum_{\chi} p(\chi) \boldsymbol\Delta_\chi.
\ee
By rearranging the remaining terms in~(\ref{ebc16.07.09-1}) one can further simplify the expression of the error matrix to obtain
\be
\boldsymbol\Delta= \mean{\Blmda\Blmda^t}-\sum_{\chi}p(\chi)\mean{\Blmda}_{\chi}\mean{\Blmda^t}_{\chi},
\label{eq:Drs}
\ee
where it is important to note that the first average is over the prior distribution $\pi(\Blmda)$ alone, i.e., independent of the measurements we may perform on the average state.
As to the second term, we may write the average value of~$\Blmda$~as
\begin{equation}
\mean{\Blmda}_{\chi}=
 \sum_{\alpha} \frac{\tilde{\boldsymbol\Lambda}_{\alpha}\,\tr E_\chi \rho_\alpha}{p(\chi)}+ \mean{\Blmda},
\end{equation}
where we have defined $\tilde{\boldsymbol\Lambda}_{\alpha}=\mean{\Blmda c_{\alpha}(\Blmda)}-\mean{\Blmda}\mean{c_{\alpha}(\Blmda)}$ and used that $\sum_{\alpha} \mean{c_{\alpha}(\Blmda)} \tr E_\chi \rho_\alpha=p(\chi)$.
Inserting this result in \eqref{eq:Drs} we find,
\be
\boldsymbol\Delta=\boldsymbol\Lambda-\sum_{\chi} \frac{\left(\sum_{\alpha} \tilde{\boldsymbol\Lambda}_{\alpha}  \tr E_\chi \rho_\alpha\right)\left(\sum_{\beta} \tilde{\boldsymbol\Lambda}{}^t_{\beta}\tr E_\chi \rho_\beta\right)}{p(\chi)},
\label{ebc17.07.09-1}
\ee
where $\boldsymbol\Lambda=\mean{\Blmda\Blmda^t}-\mean{\Blmda}\mean{\Blmda^t} $ is the covariance matrix of the~unknown weights, i.e., its elements are the second order moments of~the prior distribution $\pi(\Blmda)$.
In order to interpret the second term in this equation, we define an {\em effective} state $\sigma_\Blmda$ that combines information relative to the prior distribution of $\Blmda$ with the quantum states~$\rho_\alpha$:
\be
\sigma_{\Blmda}=\mean{\rho_{\Blmda}}+(\Blmda-\bar\Blmda)^t\left( \sum_{\alpha}\tilde{\boldsymbol\Lambda}_\alpha \rho_{\alpha}\right),\label{effsta}
\ee
where
we have defined~$\bar\Blmda=\mean{\Blmda}$.
It is shown in Appendix~\ref{effectivestate} that this equation defines a proper density matrix.  Let  $p_\Blmda(\chi)$ be the probability distribution of the outcomes obtained when performing the POVM measurement  $\{E_\chi\}$ on this effective state, namely $p_{\Blmda}(\chi)=\tr E_{\chi}\sigma_{\Blmda}$.
Then, Eq.~(\ref{ebc17.07.09-1}) can be written in a very appealing form as
\begin{equation}
\boldsymbol\Delta=\boldsymbol\Lambda-\boldsymbol F(\bar\Blmda),
\label{eq:DeltaFish}
\end{equation}
where $\boldsymbol F(\Blmda)$ is the Fisher information matrix of the probability distribution $p_{\Blmda}(\chi)$, whose elements are defined by
\be
F_{r s}(\Blmda)=\sum_{\chi} \frac{\partial_{r}p_{\Blmda}(\chi) \partial_{s}p_{\Blmda}(\chi) }{p_{\Blmda}(\chi)} ,
\label{eq:fisher}
\ee
and we use the compact notation $\partial_r=\partial/\partial\lambda_r$.
Some comments are in order.
Note that the error matrix $\boldsymbol\Delta$ has two distinct contributions:  i) the intrinsic `error' of the random variable $\Blmda$ (that one would obtained by just guessing the weights of the quantum finite mixture without performing any measurement whatsoever), which is given by the covariance matrix $\boldsymbol\Lambda$; and ii) the Fisher Information of the effective state $\sigma_\Blmda$, which represents the information gathered from the outcomes of the measurement on the average state $\rho_\Blmda$. Naturally, this information reduces the uncertainty on the actual value of $\Blmda$, which explains the minus sign in~(\ref{eq:DeltaFish}).
Despite this very natural interpretation, one might be somehow surprised to find the Fisher Information matrix in the context of Eq.~(\ref{eq:DeltaFish}). It~usually appears in connection to the  Cram\'er-Rao bound (see Sec.~\ref{CR} below), where it provides lower bounds to the MSE in estimation problems. Typically  these lower bounds are attained only in the asymptotic limit of many identical and independent copies. Note however that relation~(\ref{eq:DeltaFish}) is an exact expression.

More interestingly for our purposes here, relation~(\ref{eq:DeltaFish}) enables us to apply known results~\cite{Hel,Hol} concerning the Fisher Information. In particular, the Braunstein and Caves inequality~\cite{BC}, which states that the Fisher Information is upper bounded by the so-called Quantum Fisher Information (QFI) matrix~$\boldsymbol H(\Blmda)$. Thus,
\be
\boldsymbol\Delta
\geq \boldsymbol\Lambda-\boldsymbol H(\bar\Blmda).
\label{eq:QFb}
\ee
Before proceeding, we recall the definition of
$\boldsymbol H(\Blmda)$. Its matrix elements, which depend only on the family of states $\sigma_\Blmda$, are given by
\begin{equation}
H_{rs}(\Blmda)=\Re\;\tr\left[
L_r(\Blmda)L_s(\Blmda)\sigma_\Blmda\right] ,
\label{ebc28.07.09-4}
\end{equation}
where the matrix $L_r(\Blmda)$ is the Symmetric Logarithmic Derivative (SLD), (implicitly) defined as
\be
\frac{1}{2}\left[L_{r}(\Blmda)\sigma_{\Blmda}+\sigma_{\Blmda}L_{r}(\Blmda)\right]=\partial_{r}\sigma_{\Blmda}.
\label{eq:sld}
\ee
Although Eqs.~(\ref{ebc28.07.09-4}) and~(\ref{eq:sld}) are particularized to the case under consideration, they also apply to a general situation where $\sigma_\Blmda$ represents an arbitrary family of states, such as that defined by $\rho_\Blmda$.
We also recall that the SLD is most easily computed
in the basis that diagonalizes $\sigma_{\Blmda}$.  A simple calculation leads~to
\begin{equation}
\boldsymbol L(\Blmda)=2\sum_{n,m} \frac{\bra{\phi_{n}}\sum_{\alpha} \tilde{\boldsymbol\Lambda}_{\alpha}\rho_{\alpha}\ket{\phi_{m}}}{\nu_{m}+\nu_{n}}\ketbra{\phi_{n}}{\phi_{m}} ,
\label{ebc31.07.09-1}
\end{equation}
where $\{\ket{\phi_{n}}\}$ and $\nu_{n}$ are the eigenvectors and eigenvalues of $\sigma(\bar\Blmda)=\mean{\rho_{\Blmda}}$ respectively.

Let us go back to Eq.~(\ref{eq:QFb}). Since $\boldsymbol H(\bar\Blmda)$ is independent of the measurement (as pointed out above, it only depends on the effective state~$\sigma_\Blmda$), Eq.~(\ref{eq:QFb}) provides an absolute lower bound to the error matrix $\boldsymbol\Delta$.

In those cases where this lower bound is attainable [such as the dimension two case, where $\Blmda=(\lambda,1-\lambda)^t$, or when the SLD matrices commute with one another], the QFI matrix further provides us with the optimal measurement. In those cases $\{E_{\chi}\}$ can be chosen to be the projectors onto the
eigenspaces of $L_{r}(\bar\Blmda)$.
An important instance is the estimation of the linear combination $a=\boldsymbol a^t\Blmda$. In this case the optimal measurement is given by the projector onto the eigenspaces of $L_{a}=\sum_{r}a_{r} L_{r}(\bar\Blmda)$, and the minimal error $E_a$ is exactly given by
\begin{equation}
E_{a}= \boldsymbol a^t\boldsymbol\Lambda \boldsymbol a-2\sum_{n,m} \frac{|\bra{\phi_{m}}\sum_{\alpha} \boldsymbol a^t  \tilde{\boldsymbol\Lambda}_{\alpha}\,\rho_{\alpha}\ket{\phi_{n}}|^2}{\nu_{m}+\nu_{n}},
\label{a}
\end{equation}
which comes from sandwiching~Eq.~(\ref{eq:DeltaFish}) with $\boldsymbol a^t$ and~$\boldsymbol a$.
In particular, the MSE on a single weight $\lambda_r$ is given by
\begin{equation}
\Delta_{r r}=\Lambda_{r r}-2\sum_{n,m} \frac{|\bra{\phi_{m}}\sum_{\alpha} \tilde\Lambda_{\alpha r}\rho_{\alpha}\ket{\phi_{n}}|^2}{\nu_{m}+\nu_{n}}.
\label{eq:Drr}
\end{equation}

Quantum finite mixtures of orthogonal states ($\rho_{\alpha}\rho_{\beta}=0$) is yet another instance where the bound~(\ref{eq:QFb}) is attainable. In this case, one can easily check that the MSE is simply given by
\be
E^\perp=\tr\boldsymbol\Delta=\tr\boldsymbol\Lambda+\sum_{\alpha}\frac{\tilde{\boldsymbol\Lambda}{}^t_{\alpha}\tilde{\boldsymbol\Lambda}_{\alpha}}{\mean{c_{\alpha}(\Blmda)}}.
\label{ebc18.07.09-3}
\ee

\subsection{Estimation with multiple copies}\label{febc18.07.09-1}

Let us assume that we are given an arbitrary number $N$ of identical and independent copies of the average state $\rho_\Blmda$ in \eqref{eq:rhol}. The global state of the $N$ copies can be written~as
\be\label{rotensorn}
\rho_\Blmda^{\otimes N}=N!\sum_{\boldsymbol{k}}  \prod_{r} \frac{\lambda_{r}^{k_{r}}}{k_r!} \mathcal{S}( \rho_{1}^{\otimes k_{1}}\otimes\ldots\otimes \rho_{M}^{\otimes k_{M}}) ,
\ee
where the components of the `occupation number' vector~$\boldsymbol{k}$ satisfy $\sum_{r=1}^M k_{r}=N$, and $\mathcal{S}$ indicates averaging over all permutations of the $N$ copies, which produces a proper (normalized) state.
From this equation we note that the state $\rho_\Blmda^{\otimes N}$ can be written in the form~\eqref{eq:rhog} with $\boldsymbol k$ playing the role of $\alpha$ and $c_{\boldsymbol{k}}(\Blmda)= N! \prod_{r} \lambda_{r}^{k_{r}}/k_r!$,  $\rho_{\boldsymbol{k}}= \mathcal{S}( \rho_{1}^{\otimes k_{1}}\otimes\ldots\otimes \rho_{M}^{\otimes k_{M}})$.
Because of this, the results of the previous section can be applied to multiple copies.

For arbitrary prior distributions $\pi(\Blmda)$ that is about all we can say concerning the multiple copy scenario. However, more explicit expression can be derived if a flat distribution of weights can be assumed. This is the most conservative scenario, and also the situation when nothing is known a priori about the weights~$\Blmda$.  Appendix~\ref{form} collects useful formulae for computing integrals and averages on the simplex when~$\pi(\Blmda)$ is flat (constant). From this appendix one can easily obtain
\begin{eqnarray}
\Lambda_{r s}&=&\frac{\delta_{rs}-1/M}{M(M+1)},
\label{ebc18.07.09-2}\\[.2em]
\tilde\Lambda_{\boldsymbol{k} r}&=&\frac{k_{r}-N/M}{(N+M)}\mean{c_{\boldsymbol{k}}(\Blmda)} ,
\label{eq:lmFLAT}
\end{eqnarray}
 for the matrix elements of $\boldsymbol\Lambda$ and $\tilde{\boldsymbol\Lambda}_{\boldsymbol k}$ respectively, where
\be
\mean{c_{\boldsymbol{k}}(\Blmda)}={N+M-1 \choose N}^{-1} .
\label{ebc18.07.09-1}
\ee
Hence, the lower bound on the MSE follows:
\begin{equation}
\tr\boldsymbol\Delta\geq\frac{M-1}{M(M+1)}-2\sum_{n,m} \frac{|\bra{\phi_{m}}\sum_{\boldsymbol{k}} \tilde{\boldsymbol\Lambda}_{\boldsymbol{k}}\,\rho_{\boldsymbol{k}}\ket{\phi_{n}}|^2}{\nu_{m}+\nu_{n}}.
\end{equation}
This is as far as one can get for mixtures of arbitrary states $\{\rho_r\}$.
In the case of mixtures of orthogonal states we can substitute Eqs.~(\ref{ebc18.07.09-2}) to~(\ref{ebc18.07.09-1}) in Eq.~(\ref{ebc18.07.09-3}) and find a closed expression for the MSE for multiple copies:
\begin{equation}
E^{\perp}_{N}=\tr\boldsymbol\Delta= \frac{M-1}{(M+1)(M+N)},\label{ort}
\end{equation}
where we have used the summation formula in Appendix~\ref{form}. Note that the error $E^{\perp}_{N}$ vanishes as $N$ goes to infinity.

\subsection{Identifiability}{\label{sec:identifiability}}

A mixture is
{\em identifiable} if there exists a one-to-one correspondence between $\Blmda$ and $\rho_\Blmda$. That is
to say, iff
given $\rho_\Blmda$, there is no other vector of weights $\Blmda$ satisfying~Eq.~(\ref{eq:rhol}).
 In a general situation, though,  different vectors~$\Blmda$ can give rise to the same density matrix ($\rho_\Blmda=\rho_{\Blmda'}$ for some $\Blmda\not=\Blmda'$)  and, therefore, identifiability
cannot be taken for granted.
Necessary and sufficient conditions for identifiability of classical finite mixtures, $p_{\Blmda}(i)= \sum\lambda_{r}p_{r}(i)$, were established more than four decades ago by Teicher \cite{teich1963}. These conditions  are equivalent to $\{p_{r}\}_{r=1}^M$ being a linearly independent set.
Similarly, the linear independence of the (density) matrixes in a quantum ensemble~$\{\rho_{r}\}_{r=1}^M$, constitutes a necessary and sufficient condition for the identifiability of quantum finite mixtures: states lying in the convex hull of a linearly independent set of density matrixes will be identifiable, while all states in the convex hull of a linearly dependent set will necessarily be unidentifiable, except for possibly some states on the boundary.

Identifiability is usually assumed in (classical) mixture estimation  (see e.~g.~\cite{Boas}), since unidentifiable models often give rise to ill-defined estimation procedures
and their asymptotic theories break down. In contrast,  our approach leads to sensible results for the estimation of quantum finite mixtures {\em even in unidentifiable scenarios}.
The above results for single-copy case, as well as the derivation of the effective model for finite number of copies, can be directly applied without taking notice of identifiability considerations. Care must be taken, however, when applying the asymptotic methods of the next section to unidentifiable mixtures. Such methods assume that the errors go to zero as the number of copies increases,   which cannot be guaranteed if mixtures are unidentifiable. We will revisit unidentifiability at the end of Sec.~\ref{sec3}, where we will introduce ways to circumvent this difficulty.

\section{Estimation of weights in the asymptotic limit}\label{sec3}

In the preceding sections  we have presented protocols to optimally estimate quantum finite mixtures and have obtained bounds on their accuracy using a covariance-type error matrix as a cost functions. We have also identified situations where these bounds are attainable and provided the corresponding optimal measurements; all this, in the framework of single- and multiple-copy estimation. In this section, we focus on the latter, in the asymptotic limit when a large number~$N$ of copies is available for the experiment.  Although the approach of the preceding sections can be carried out  also in this case, asymptotic expansions become involved, with a few exceptions where a closed expression can be found for arbitrary~$N$  [see, e.g., Eq.~(\ref{ort})]. Our aim here is to provide more straightforward means to obtain asymptotically optimal estimation protocols in general situations and compute the corresponding  MSE.
For this, we can resort on the well known Cram\'er-Rao (CR) theory and its quantum extension, which we briefly discuss next, particularized to finite mixture estimation. A very powerful result, known as Holevo bound, will be also presented in the next section along with a simple example of use. A more detailed and comprehensive presentation, which includes a discussion on the relationship between this theory and the Bayesian approach of the preceding sections, can be found in~\cite{Bag}.

In this framework, to which we will refer as `pointwise', one focusses on a fix point in parameter space, i.e., the unit simplex in our case, and restrict oneself to consider Locally Unbiassed (LU) estimators: those for which  $\langle\Blmda_\chi\rangle_\Blmda=\Blmda$ in some open set, where, in the same spirit of previous notation, $\langle\;\cdot\;\rangle_\Blmda$ indicates averaging over the conditional probability $p(\chi|\Blmda)$ at the fixed point~$\Blmda$. We define the error matrix $\boldsymbol\Delta(\Blmda)$ as
\begin{eqnarray}
\boldsymbol\Delta(\Blmda)&=&\langle(\Blmda_\chi-\Blmda)(\Blmda_\chi-\Blmda)^t\rangle_\Blmda
\nonumber\\
&=&
\sum_\chi p(\chi|\Blmda) (\Blmda-\Blmda_\chi)(\Blmda-\Blmda_\chi)^t
.
\end{eqnarray}
It depends on the measurement and on the estimator, i.e., on the particular way one associates $\Blmda_\chi$ to a given outcome~$\chi$ of the chosen measurement.
For the sake of simplicity in most of  this section we will assume that the mixtures are identifiable. The problem of dealing with unidentifiable mixtures will be postponed to the last subsection~(\ref{subsecUnId}).

\subsection{The Cram\'er Rao Bound}\label{CR}

A first important result of the theory is the so-called CR bound~\cite{cramer,rao}. It states that the error matrix of a LU estimator
at~$\Blmda$ is lower bounded by the inverse of  the  Fisher Information defined in~(\ref{eq:fisher}) with $p_\lambda(\chi)=p(\chi|\Blmda)=\tr\,E_{\chi,\Blmda}\,\rho_\Blmda$ [note that in this theory the POVM may depend on the vector~$\Blmda$; see the comments after Eq.~(\ref{ebc28.07.09-3})],
namely,
\begin{equation}
\boldsymbol\Delta(\Blmda)\geq \boldsymbol F^{-1}(\Blmda).
\end{equation}
Assume now that the same measurement is performed on several independent copies; i.e., on the average state~$\rho_\Blmda^{\otimes N}$.
Due to the additivity of the Fisher Information, in this multiple-copy scenario
one has
\begin{equation}
\boldsymbol\Delta(\Blmda)\geq {\boldsymbol F_1^{-1}(\Blmda)\over N},
\label{ebc28.07.09-2}
\end{equation}
where the subscript~$1$ refers to the one-copy model $\rho_\Blmda$. This inequality expresses the fact that the MSE of the estimation scales with the inverse of the number of copies, and the accuracy by which we are able to estimate $\Blmda$ with just a copy sets the scale.
It is well-known that under some
regularity conditions the maximum likelihood  estimator achieves the
CR bound asymptotically.

In spite of  its fundamental character, the CR bound has the drawback that the bound it provides refers to
a particular measurement, not necessarily optimal. To go around this difficulty, we invoke the Braunstein and Caves inequality, already discussed in the Sec.~\ref{ebc28.07.9S-1},
and obtain
\be
\boldsymbol\Delta(\Blmda)\geq\frac{\boldsymbol H_1^{-1}(\Blmda)}{N}.
\label{ebc28.07.09-3}
\ee
Recall, however, that this bound is not always attainable but, when it is, the projectors onto the eigenspaces of the SLD $L_r(\Blmda)$ define the optimal measurement.
It is important to point out here that practical use of this approach requires a two-step measurement in order to saturate the bound. This is necessary because this optimal measurement, and thus the estimator, depend themselves on $\Blmda$, which we do not know beforehand. To overcome this difficulty, one can take an asymptotically vanishing fraction
of copies, say $\sqrt{N}$, and make an initial estimate of the
weights $\Blmda_{\rm ini}$. Then, on the remaining copies one can
perform the measurement that is optimal at $\Blmda_{\rm ini}$, i.e., project on the eigenspaces of~$L_r(\Blmda_{\rm ini})$ (see Appendix \ref{two-step} for an explicit example of this procedure). Thus, this two-step adaptive measurement, which is {\em independent} of~$\Blmda$, approaches the optimal one in the asymptotic limit at leading order in $1/N$, and one may write
\begin{equation}
\boldsymbol\Delta=\int d\Blmda\;\pi(\Blmda) \boldsymbol\Delta(\Blmda)+o\left(N^{-1}\right) .
\end{equation}
This equation establishes a bridge between the asymptotic pointwise theory of this section and the Bayesian approach discussed in the first part of this paper.
With all this in mind, we conclude that for sufficiently smooth priors $\pi(\Blmda)$ it holds that
\be
\boldsymbol\Delta \geq\frac{1}{N}\int d\Blmda\;\pi(\Blmda) \boldsymbol H_1^{-1}(\Blmda)+o\left(N^{-1}\right) .
\label{ebc28.07.09-1}
\ee

So far in this section we have overlooked the fact that not all the components of~$\Blmda$ are independent, as $\Blmda$ must lie on the unit simplex. One could circumvent this by simply using the constraint $\sum_r\lambda_r=1$ to write a particular component, say $\lambda_M$, in terms of the remaining~$M-1$ as $\lambda_M=1-\sum_{r=1}^{M-1}\lambda_r$. This possibility, however, introduces a huge asymmetry in the calculation which may result in difficulties to invert the Fisher Information matrix $\boldsymbol H_1$ and compute the bound~(\ref{ebc28.07.09-1}). Note that inside the unit simplex the variations of $\Blmda$ are constrained by~$\Delta\Blmda\cdot \boldsymbol u=0$, where $\boldsymbol u=(1,1,\dots,1)$. A fully symmetric way of dealing with this issue is to project the information matrices $\boldsymbol F$ and $\boldsymbol H$ onto the orthogonal complement of span$\{\boldsymbol u\}$, which we call~$S$.
Thus, the CR bound, Eq.~(\ref{ebc28.07.09-2}), takes the form~\cite{gorman}
\be
{\boldsymbol P}_S\boldsymbol\Delta(\Blmda){\boldsymbol P}_S\geq\frac{1}{N}\left[ {\boldsymbol P}_S {\boldsymbol F}_1(\Blmda) {\boldsymbol P}_S\right]^{-1},
\ee
and similarly for its quantum version in Eq.~(\ref{ebc28.07.09-3}),
where~${\boldsymbol P}_S$ stands for the projector on~$S$ and the inverse,~$[\;\cdot\;]^{-1}$,  is restricted to the support of~${\boldsymbol P}_S$.

As an example, let us consider again the mixture of~$M$ orthogonal states and compute the asymptotic expression of $E^{\perp}_{N}$, introduced in~(\ref{ort}). Applying the definition of SLD in Eq.~(\ref{eq:sld}) to the 1-copy family $\rho_\Blmda$ it is straightforward to obtain that $L_r(\Blmda)=P_r/\lambda_r$, where $P_r$ is the projector onto the support of~$\rho_r$. Applying now the definition of the QFI, Eq.(\ref{ebc28.07.09-4}), to the same family we obtain $[H_1(\Blmda)]_{rs}=\delta_{rs}/\la_r$. For brevity, we omit the arguments and write $\boldsymbol H_1'$ for the projection of~$\boldsymbol H_1$ onto~$S$, i.e., $\boldsymbol H_1'={\boldsymbol P}_S\boldsymbol H_1{\boldsymbol P}_S$, and similarly for other matrices. Let us start by computing $\det \boldsymbol H_1'$ (here the zero eigenvalue corresponding to the kernel of the projection is, of course, removed from $\det$). %For a diagonal matrix of the form $M_{rs}=\mu_r \delta_{rs}$.
Since i)~the determinant~of a~$d\times d$ matrix is a homogeneous polynomial of degree~$d$ in its matrix elements and ii)~the vector~$\boldsymbol u$ has the same projection on each eigenspace of $\boldsymbol H_1$, it follows that  i)~$\det \boldsymbol H_1'$ must also be a homogeneous polynomial of degree $M-1$ in $1/\lambda_r$, i.e., in the eigenvalues of~$\boldsymbol H_1$, and ii)~it must be a symmetric function of these eigenvalues. We also note that $\det\boldsymbol H_1'$ must vanish if any two or more of these eigenvalues are set equal to zero, since in this case~$S$ necessarily contains a null subspace of $\boldsymbol H_1$ [in doing so,  the condition $\lambda_r\le1$ is temporarily lifted, which is legitimate, since the result we are after, Eq.~(\ref{ebc02.06.09-1}) below, is an algebraic relation that holds for generic~$\{\lambda_r\}$ regardless whether they are probabilities or not]. Hence,
\begin{equation}
\det \boldsymbol H_1'={1\over M} \sum_{\{r_i\}} \prod_ {i=1}^{M-1}{1\over\la_{r_i}} ={1\over M}\prod_ {i=1}^{M}{1\over\la_{i}} ,
\label{ebc02.06.09-1}
\end{equation}
where the sum extends to all subsets of $M-1$ indexes drawn from~$1,2,\dots,M$, and the  prefactor~$1/M$ can be easily computed by considering the particular case where all $\lambda_r$ are equal. Reasoning along the same lines, we conclude that
\be
\det{\boldsymbol H'_1} \,\tr({\boldsymbol H}'_1)^{-1}={2\over M}  \sum_{\{r_j\}}\prod_{i=1} ^{M-2}{1\over\lambda_{r_i}}.
\label{ebc22.12.09-1}
\ee
[Note that on the left hand side of this last equation $\det{\boldsymbol H'_1} \,[({\boldsymbol H}'_1)^{-1}]_{st}$ is the $(s,t)$ cofactor of $\boldsymbol{H}'_1$, i.e., the signed determinant of the matrix ${\boldsymbol H}'_1$ with row $s$ and column $t$ removed. It follows that  $\det{\boldsymbol H'_1} \,\tr({\boldsymbol H}'_1)^{-1}$ is a homogeneous and symmetric polynomial in $1/\lambda_r$ of degree $M-2$ that vanishes if three or more eigenvalues of ${\boldsymbol H_1}$ are set equal to zero.]
Combining Eq.~(\ref{ebc22.12.09-1}) with Eq.~(\ref{ebc02.06.09-1}), and after some algebra, we obtain
\be
\tr({\boldsymbol H}'_1)^{-1}=\left(\sum_r \la_r\right)^{-1} \hspace{-.5em}\sum_{r\neq s} \la_{r} \la_{s}
=1-\sum_{r} \la^2_{r}.
\ee
The averaging over the flat prior can be easily performed with the help of Appendix~\ref{aver}, obtaining
\be
\int d\Blmda\;\pi_{\rm flat} \;\tr({\boldsymbol H}'_1)^{-1}={M-1\over M+1} .
\ee
Taking into account~(\ref{ebc28.07.09-1}) and $E^\perp_{N}=\tr\boldsymbol\Delta=\tr\boldsymbol\Delta'$ up~to~$o(N^{-1})$, we finally find that $E^\perp_{N}=(M-1)/[(M+1)N]+o(N^{-1})$, which indeed agrees with~(\ref{ort}) for large~$N$.

\subsection{Holevo bound}\label{HB}

The quantum CR bound is a matrix inequality which is in general non-attainable [a few remarkable exceptions are those discussed in Sec.~\ref{ebc28.07.9S-1}, in the paragraph after Eq.~(\ref{ebc31.07.09-1}), and the example above].
However there is a related bound that one can expect to be
saturated asymptotically: the Holevo bound.   Indeed for qubit systems asymptotic attainability has been proved by Hayashi and Matsumoto in ~\cite{M&K} and the general proof for finite dimensional systems follows from a recent paper by Kahn and Gu\c{t}\u{a} \cite{kahn2009}. We note that attainability here, as in the CR bound, is proven in a pointwise approach and hence makes implicit use of  the two step adaptive measurement that we mentioned above. An important difference here is that at the second step the measurement attaining the Holevo bound will in general be a collective measurement that can not be implemented by local measurements on each copy.

Let us briefly introduce the Holevo bound for quantum finite mixture estimation (see also~\cite{Bag}).
Let $G$ be a positive semi-definite matrix
and
\begin{equation}\label{def of C^N}
C^N_{\Blmda}(G) =\min_{\mbox{\tiny $\displaystyle
\begin{array}{c}
\{(\{E_\chi\}, \{\Blmda_\chi\})\}\\
\mbox{LU}
\end{array}
 $}}
\tr\, G
\boldsymbol\Delta(\Blmda),
\end{equation}
where the minimization is over all pairs $(\{E_\chi\},\{\Blmda_\chi\})$ of measurements on~$\rho^{\otimes N}_\Blmda$ and estimators for which the latter is LU at $\Blmda$ (the unbiasedness of an estimator depends on the measurement through its outcome probability distribution).
Eq.~(\ref{def of C^N}) is relevant to the problem we are dealing with because its right hand side
gives, e.g.,  the smallest MSE,~$\tr\,\boldsymbol\Delta(\Blmda)$, if $G=\openone$. I.e., $C^N_{\Blmda}(\openone)$ is the MSE of the optimal $N$-copy estimation scheme.

In Ref.~{\cite{Hol}} Holevo  proved the following bound:
\be
    \label{eq:holevobound}
    C^1_{\Blmda}(G) \geq C^H_{\Blmda}(G),
\ee
where
\begin{eqnarray}\label{C^H}
    C^H_{\Blmda}(G)&=&\min_{\xt\in \Xi_{\Blmda}}
  \bigg\{  \tr\, G \,\Re Z[\xt]\nonumber\\
    &+& \tr \left| \sqrt{G} \,\Im Z[\xt]\sqrt{G}\,\right|  \bigg\}.
\end{eqnarray}
In this expression $\xt=(X_1,X_2,\dots, X_{M-1})$ are hermitian  matrices, one for each {\em independent} parameter (Thus, for quantum mixtures, we will choose $\lambda_M=1-\sum_r^{M-1}\lambda_r$), satisfying the following relations
\begin{eqnarray}
\tr\, \rho_\Blmda \xt
    &=&0\label{eq:Xcond1},\\
\tr\, \partial_r \rho_\Blmda\, X_s &=&\delta_{rs},\quad 1\le r,s\le M-1.
\label{eq:Xcond2}
\end{eqnarray}
The minimization in~(\ref{C^H}) is over the set $\Xi_{\Blmda}$ of all such~$\xt$.
Finally,  $Z[\xt]$  is the matrix whose elements are given by
\begin{equation}
Z_{rs}  [ \xt   ]
= \tr \rho_\Blmda X_r X_s;\quad 1\le r,s\le M-1  .
\end{equation}

Although the Holevo bound~(\ref{eq:holevobound}) is not in general attainable, it is  attainable for the class of Gaussian models.
The recent work \cite{kahn2009} on asymptotic normality shows the asymptotic (local) equivalence between the many-copy states of finite-dimensional systems
and a Gaussian model and thereby proves the asymptotic attainability of the Holevo bound for finite dimensional systems,
  i.e.,%
\be \label{eq:asympattholevobound}
\lim_{N\to\infty}N C^N_{\Blmda}(G) =C^H_{\Blmda}(G),
 \ee

To relate the above with the Bayesian approach of the preceding sections, we need to average over $\pi(\Blmda)$: asymptotically, we have $\tr\,G\boldsymbol\Delta=N^{-1}\int d\Blmda\, \pi(\Blmda)\, C^H_\Blmda(G)+o(N^{-1})$. Thus, for instance, the MSE~$E$ can be computed as
\begin{equation}
E=\tr\,\boldsymbol\Delta={1\over N}\int d\Blmda\,\pi(\Blmda) C^H_\Blmda(\openone)+o\left(N^{-1}\right)  .
\label{ebc01.08.09-1}
\end{equation}
As to whether or not this averaging is legitimate and the resulting bound on the averaged cost function is attainable, there exist very good heuristic arguments,
as well as various examples~\cite{Bag}, that this should be the case,  but no rigorous proof. Thus, this last equation should be taken with a grain of salt.

To illustrate the use of the Holevo bound in finite mixture estimation, let us assume that $\rho_r$, $1\le r\le4$ are four pure qubit states whose Bloch vectors~$\vec n_r$  form the vertices of a regular tetrahedron:
\begin{equation}
\begin{array}{rclcrcl}
\vec n_1&=&\displaystyle{1\over\sqrt3}
\begin{pmatrix}
\phantom{-}1\cr-1\cr-1
\end{pmatrix},
&\quad&
\vec n_2&=&\displaystyle{1\over\sqrt3}
\begin{pmatrix}
-1\cr\phantom{-}1\cr-1
\end{pmatrix},\\[2.em]
\vec n_3&=&\displaystyle{1\over\sqrt3}
\begin{pmatrix}
-1\cr-1\cr\phantom{-}1
\end{pmatrix},&&
\vec n_4&=&\displaystyle{1\over\sqrt3}
\begin{pmatrix}
1\cr1\cr1
\end{pmatrix} .
\end{array}
\end{equation}
With this, the Bloch vector of the finite mixture is $\vec r_\Blmda=\vec n_4+\sum_{r=1}^3\lambda_r(\vec n_r-\vec n_4)$. With full generality we may write $X_r=a_r+\vec b_r\cdot\vec\sigma$, where $\vec\sigma=(\sigma_x,\sigma_y,\sigma_z)$ are the standard Pauli matrixes. Conditions~(\ref{eq:Xcond1}) and~(\ref{eq:Xcond2}) are equivalent to $a_r=-\vec b_r\cdot\vec r_\Blmda$ and $(\vec n_r-\vec n_4)\cdot\vec b_s=\delta_{rs}$, $1\le r,s\le 3$. This very last equation can be inverted (this will be always the case if the mixture is identifiable)  and we obtain
$\vec b_r=(3/4)\vec n_r$, $1\le r,s\le 3$.
With this,
\begin{equation}
X_r={1-4\lambda_r+3\vec n_r\cdot\vec\sigma\over4},\quad r=1,2,3.
\end{equation}
We see that~(\ref{eq:Xcond1}) and~(\ref{eq:Xcond2}) determine $\boldsymbol X$ uniquely and no minimization is required in~(\ref{C^H}). A straightforward calculation leads to
\begin{eqnarray}
(\Re\,Z[\boldsymbol X])_{rr}&=&{(1+2\la_r)(1-\la_r)\over2};\\
(\Re\,Z[\boldsymbol X])_{rs}&=&-{3+(1-4\la_r)(1-4\la_s)\over16},\; r\not=s;\\
(\Im\,Z[\boldsymbol X])_{rs}&=&{\sqrt3\over4}\epsilon_{rst}(\la_4-\la_t);
\end{eqnarray}
where $\epsilon_{rst}$ is the (fully antisymmetric) Levi-Civita tensor in three dimensions. To compute the MSE we need the following
\begin{eqnarray}
\tr\,\Re\,Z[\boldsymbol X]&=&{1\over2}\left[3+\sum_{r=1}^3 \la_r(1-2\la_r)\right],
\label{ebc06.10.09-1}
\\[.2em]
\tr\left|\Im\,Z[\boldsymbol X]\right|&=&
{\sqrt3\over2}\left[{\sum_{r=1}^3(\la_4-\la_r)^2}\right]^{1/2},
\label{ebc06.10.09-2}
\end{eqnarray}
and, averaging over $\pi_{\rm flat}(\Blmda)$, Eq.~(\ref{ebc01.08.09-1}):
\begin{equation}
E={1\over N}\left({63\over40}+0.43\right)+o\left(N^{-1}\right)
={2.01\over N}+o\left(N^{-1}\right)
\label{ebc01.08.09-2}
\end{equation}
where the first (second) figure in the parenthesis comes from the real (imaginary) part of $Z[\boldsymbol{X}]$ in~Eq.~(\ref{ebc06.10.09-1}) [Eq.~(\ref{ebc06.10.09-2})].
It is interesting to note that for this example, the quantum CR bound is not attainable. Indeed, one can check that the SLD $L_r(\Blmda)$ is given by
\begin{equation}
L_r(\Blmda)={\vec n_r\cdot \vec r_\Blmda\over|\vec r_\Blmda|^2-1}+\left(
\vec n_r-{\vec n_r\cdot \vec r_\Blmda\over|\vec r_\Blmda|^2-1}
\,\vec r_\Blmda
\right)\cdot\vec\sigma ,
\end{equation}
where $\vec r_\Blmda=\sum_{r=1}^4\la_r\vec n_r$ is the Bloch vector of the averaged state $\rho_\Blmda$ and $1\le r\le 4$ (we now treat all components of~$\Blmda$ as independent, in accordance with the approach developed in~Sec.~\ref{CR}). One can immediately check that the commutator of the SLDs does not vanish, and the quantum CR bound is not saturated. Just for the sake of completeness, the quantum Fisher Information matrix is given by
\begin{equation}
(H_1)_{rs}=\vec n_r\cdot\vec n_s+{(\vec n_r\cdot \vec r_\Blmda)
(\vec n_s\cdot \vec r_\Blmda)\over1-|\vec r_\Blmda|^2},
\end{equation}
for $1\le r\le4$. Projecting on~$S$ with
\begin{equation}
\boldsymbol P_S=
\begin{pmatrix}
-1&-1&-1\cr
\phantom{-}1&\phantom{-}0&\phantom{-}0\cr
\phantom{-}0&\phantom{-}1&\phantom{-}0\cr
\phantom{-}0&\phantom{-}0&\phantom{-}1
\end{pmatrix} ,
\end{equation}
and (pseudo)-inverting, one obtains the relation
\begin{equation}
\boldsymbol H_1^{-1}=\Re\, Z[\boldsymbol X].
\end{equation}
After averaging, we observe from~(\ref{ebc01.08.09-2}) that the quantum~CR bound
\begin{equation}
E> {63\over40N}+o\left(N^{-1}\right)
\end{equation}
cannot be saturated.

\subsection{Unidentifiable mixtures in the asymptotic limit}\label{subsecUnId}

In the preceding section, we were required to assume  that estimation errors become vanishingly small as the number of copies increases. This assumption does not necessarily  hold if mixtures are unidentifiable.
In order to be able to apply the asymptotic techniques introduced above, we make a useful observation. If a quantum finite mixture is unidentifiable there necessarily exists an orthogonal transformation  $\Blmda'=\boldsymbol{\mathscr O}\Blmda$ such that the states~$\rho_{\Blmda}$ depend solely on a reduced number of parameters $\{\xi_r=\lambda_{r}'\}_{r=1}^m$, with~$m<M$, and are independent of  the redundant parameters $\{\eta_r=\lambda'_{r}\}_{r=m+1}^M$. The error matrix~$\boldsymbol{\Delta}$ of the original parameters $\Blmda$ is, of course, related to the error matrix $\boldsymbol{\Delta}'$  of the new ones,~$\boldsymbol{\xi}$,~$\boldsymbol{\eta}$,  by the similarity transformation~$\boldsymbol{\Delta}'=\boldsymbol{\mathscr O} \boldsymbol{\Delta} \boldsymbol{\mathscr O}^t$. Any measurement performed on the state $\rho_{\Blmda}$ will only give information about the parameters~$\boldsymbol{\xi}$,
whereas the components of~$\boldsymbol{\eta}$  have to be
guessed independently of the measurement outcomes (e.g., by random choice). The optimal choice for $\boldsymbol\eta$ is, actually, $\langle\boldsymbol{\eta}\rangle$, and leads to an error that is, of course,  independent of the number of copies.
 This means that in unidentifiable quantum mixture estimation there will always be an intrinsic error associated to the uncertainty in the redundant parameters $\boldsymbol\eta$, which remains constant regardless of the number of copies one is provided with. In the asymptotic limit, one can apply the bounds of the preceding sections to the block of~$\boldsymbol{\Delta}'$ corresponding to the relevant components~$\boldsymbol{\xi}$.

To illustrate this let us consider the unidentifiable qubit mixture defined by
\begin{eqnarray}
\rho_\Blmda&=&\lambda_1|0\rangle\langle0|+ \lambda_2|1\rangle\langle1|+ \lambda_3|+\rangle\langle+|+ \lambda_4|-\rangle\langle-|\nonumber\\
&=&\frac{1}{2}\left[\openone+
(\lambda_1-\lambda_2)\sigma_z+(\lambda_3-\lambda_4)\sigma_x
             \right],
\end{eqnarray}
where $|\pm\rangle=(|0\rangle\pm|1\rangle)/\sqrt2$.
If we perform the following rotation $\boldsymbol{\mathscr O}$ in parameter space
\be
\begin{pmatrix}
\xi_1\\ \xi_2 \\ \eta_1 \\ \eta_2
\end{pmatrix}
=\frac{1}{\sqrt{2}}\left(
                               \begin{array}{rrrr}
                                 1 & -1 & 0 & 0 \\
                                 0 & 0 & 1 & -1 \\
                                 1 & 1 & 0 & 0 \\
                                 0 & 0 & 1 & 1 \\
                               \end{array}
                             \right)
\begin{pmatrix}
\lambda_1\\ \lambda_2 \\ \lambda_3 \\ \lambda_4
\end{pmatrix}
,
\ee
we have
\begin{equation}
\rho_{\Blmda}={1\over2}\left(\openone+\sqrt2\,\xi_1\sigma_z+\sqrt2\,\xi_2\sigma_x\right).
\label{ebc01.10.09-1}
\end{equation}
This shows that~$\eta_1$ and~$\eta_2$ are redundant parameters, and measurements will give us no information about them. For the simple model in~(\ref{ebc01.10.09-1}), it is straightforward to obtain the Holevo bound. We first check that~$\boldsymbol{X}=(X_1,X_2)^t$, with
\begin{equation}
X_1=
{\sigma_z\over\sqrt2}-\xi_1\,\openone,\qquad
X_2={\sigma_x\over\sqrt2}-\xi_2\,\openone ,
\end{equation}
is the solution to conditions~(\ref{eq:Xcond1}) and~(\ref{eq:Xcond2}) that minimizes~(\ref{C^H}). It follows that
\begin{equation}
\label{Z[X]unident}
\Re\, Z[\boldsymbol X]=
\begin{pmatrix}
{1\over2}-\xi_1^2 && -\xi_1\xi_2 \\[.5em]
-\xi_1\xi_2 && {1\over2}-\xi_2^2
\end{pmatrix},
\quad
\Im\, Z[\boldsymbol X]=0,
\end{equation}
and~(\ref{C^H}) gives
\begin{equation}
C^H_{\boldsymbol{\xi}}(\openone)=1-\xi_1^2-\xi_2^2 .
\end{equation}
In the limit $N\to\infty$,
we can compute the error coming from the estimation of~$\boldsymbol\xi$ through~(\ref{ebc01.08.09-1}), i.e.,
\begin{eqnarray}
\sum_{r=1}^2\Delta^{(\boldsymbol\xi)}_{rr}\!\!&=&\!\!
{1\over N}\!\!\int \!\!d\Blmda\,\pi_{\rm flat}(\Blmda) \,C^H_{\boldsymbol{\xi}(\Blmda)}(\openone)+o\left(N^{-1}\right)
\nonumber\\
\!\!&=&\!\!{9\over10N}+o\left(N^{-1}\right)  ,
\end{eqnarray}
which is asymptotically vanishing.
As for the estimation of~$\eta_1$ and~$\eta_2$, we  make the optimal guess
\begin{equation}
\langle\eta_r\rangle=\int d\Blmda\,\pi_{\rm flat}(\Blmda)\, \eta_r(\Blmda)={1\over2\sqrt2} ,
\end{equation}
 thus
\begin{equation}
\sum_{r=1}^2\Delta^{(\boldsymbol\eta)}_{rr}\!=\!\!
\int \!\!d\Blmda\,\pi_{\rm flat}(\Blmda) \!
\sum_{r=1}^2\left[\eta_r(\Blmda)-\langle\eta_r\rangle\right]^2
={1\over20}
\end{equation}
[according to the notation introduced in the paragraph below~(\ref{ebc17.07.09-1}), this quantity could also be denoted by~$\sum_{r=1}^2\Lambda^{(\boldsymbol\eta)}_{rr}$].

Putting all pieces together, the estimation error is
\begin{eqnarray}
E&=&\tr\,\boldsymbol{\Delta}=\tr\,\boldsymbol{\Delta}'
=\sum_{r=1}^2\Delta^{(\boldsymbol{\eta})}_{rr}+
\sum_{r=1}^2\Delta^{(\boldsymbol{\xi})}_{rr}\nonumber
\\
&=&
{1\over20}+{9\over10N}  .
\end{eqnarray}
In conclusion, this explicit example shows that unidentifiable mixtures will lead to a non-vanishing estimation error even in the asymptotic limit.

\section{Estimation of two-component mixtures}\label{sec4}

In this section we dwell on the simplest quantum mixture scenario, where the average state~$\rho_\Blmda$ belongs to the $1$-simplex
\begin{equation}\label{problem}
\rho_\lambda=\lambda\rho_1+(1-\lambda)\rho_2,  \quad  0\le\lambda\le1
\end{equation}
(hence $\lambda_1=\lambda$, $\lambda_2=1-\lambda$).
Although the error matrix is $2\times2$, only one of its entries, say $\Delta_{11}$, contains independent information about the accuracy in the estimation of the mixture~(\ref{problem}). Therefore, in the following we simply drop the remaining three entries and write~$\Delta$ [and likewise for the Fisher information matrix~$\boldsymbol{F}(\Blmda)$, the quantum Fisher~$\boldsymbol{H}(\Blmda)$, etc., to which we will refer as~ $F(\lambda)$,~$H(\lambda)$,~etc.].
Since there is only an independent parameter, we may also drop the vector notation  and write $\lambda$ instead of~$\Blmda$.

\subsection{Single-shot estimation}

The single-copy version of this problem was considered recently in~\cite{Kon}, though the optimal measurements and minimal estimation error were only determined when~$\rho_1$ and~$\rho_2$ are qubit and/or pure states. Our results in~Sec.~\ref{sec2} show that for the two-component mixture in~(\ref{problem}),  % \eqref{eq:QFb} and, hence,
the attainability conditions are fulfilled for {\em any}~$\rho_1$ and $\rho_2$,  and the optimal measurements, along with their minimal estimation error, can always be determined in both single- and multiple-copy scenarios. In particular, it follows from our results that the optimal protocol consists of a projective measurement, where the projectors are those onto the eigenspaces of the SLD~$L(\bar\lambda)$. Our results in the present paper thus provide answers to various open questions posed in~\cite{Kon}.

Let us focus first on the single-copy estimation.
By~choosing the optimal estimator~(\ref{eq:optguess}), the MSE is given by~(\ref{eq:Drr}). For the mixture~(\ref{problem}) this equation can be cast as
\be\label{ebc03.10.09-1}
\Delta = \Lambda-2\Lambda^2\sum_{nm}\frac{|\langle\phi_m|\rho_1-\rho_2|\phi_n\rangle|^2}{\nu_m+\nu_n}.
\ee
[In the case under consideration here, $\tilde\Lambda_{\alpha 1}=\langle\lambda\lambda_\alpha\rangle-\langle\lambda\rangle\langle\lambda_\alpha\rangle$. Thus, $\tilde\Lambda_{1 1}=\langle\lambda^2\rangle-\langle\lambda\rangle^2\equiv\Lambda$, and~$
\tilde\Lambda_{2 1}=\langle\lambda(1-\lambda)\rangle-\langle\lambda\rangle\langle1-\lambda\rangle=-\Lambda$.]
The bound~(\ref{ebc03.10.09-1}) is attained with the measurement characterized by the eigenprojectors of the~SLD~[see Eq.~(\ref{ebc31.07.09-1})]
\be
L(\bar\lambda)=2\Lambda\sum_{nm}\frac{\langle\phi_m|\rho_1-\rho_2|\phi_n\rangle}{\nu_m+\nu_n}|\phi_n\rangle\langle\phi_m|,
\ee
where we recall that~$\{\ket{\phi_{n}}\}$ ($\nu_{n}$) are the eigenvectors (eigenvalues) of~$\mean{\rho_\la}$. One can readily check by explicitly solving~\eqref{eq:sld} that for a uniform prior $\pi_{\rm flat}(\lambda)=1$, and for pure (or for qubit states with the same purity)~$\rho_1$ and~$\rho_2$ we have  $L(\bar\lambda=1/2)\propto(\rho_1-\rho_2)$, in agreement with~\cite{Kon}. Accordingly, in this situation $\Delta=[2+\tr(\rho_1\rho_2)]/36$.

\subsection{Multiple-copy estimation and the asymptotic limit}

Although a straightforward exercise, computing $\Delta$  for~$N>1$ copies of~$\rho_\lambda$ is a tedious task even for two-component  mixtures. In most cases, the resulting expressions cannot be written in closed form for arbitrary~$N$ and are thus not very revealing. So, rather than attempting to present a general case, we have selected a particular example, which we will later use to illustrate the connection between the Bayesian and the asymptotic pointwise approaches.

Assume $\rho_1$ and $\rho_2$ are commuting non-orthogonal qubit states. Let us further assume that $\rho_2$ is pure and that the prior is flat. Then, we can choose basis so that
\be
\label{ebc04.10.09-2}
\rho_\lambda=\lambda \left(
                       \begin{array}{cc}
                         1-\epsilon  & 0 \\
                         0 & \epsilon  \\
                       \end{array}
                     \right)
+(1-\lambda)  \left(\begin{array}{cc}
                1 & 0 \\
                0 & 0 \\
              \end{array}
            \right).
\ee
Proceeding as in~Sec.~\ref{febc18.07.09-1}, the $N$-copy state $\rho_\lambda^{\otimes N}$ %
can be cast in the form~(\ref{eq:rhog}) with
\begin{equation}
c_k(\lambda)={N\choose k}
(\lambda\epsilon)^k(1-\lambda\epsilon)^{N-k}, \quad 0\le k\le N,
\end{equation}
and $\rho_k=\mathcal{S}[|1\rangle\langle1|^{\otimes k}\otimes|0\rangle\langle0|^{\otimes (N-k)}]$. Hence, using Eq.~(\ref{eq:Drr}) we have that the minimum error is given by [recall that we are assuming the flat prior $\pi_{\rm flat}(\lambda)=1$]
\be
\Delta^{\rm comm}=\frac{1}{12}-\frac{1}{2} \sum_{n,m} \frac{|\bra{\phi_{m}}\sum_{k}B_{k}\rho_{k}\ket{\phi_{n}}|^2}{\nu_{m}+\nu_{n}},
\ee
where
\begin{align}
B_k&\equiv2\tilde\Lambda_{k}=2\left[\langle\lambda c_k(\lambda)\rangle-\langle\lambda\rangle\langle c_k(\lambda)\rangle\right]
\nonumber\\
&={N \choose k} \int_0^1 d\lambda \,(2\lambda-1)(\epsilon\lambda)^k (1-\epsilon\lambda)^{N-k},
\label{ebc01.10.09-2}
\end{align}
and where now
\begin{equation}
\{\ket{\phi_{n}}\}={\rm perms}\left\{|1\rangle^{\otimes k}\otimes|0\rangle^{\otimes (N-k)}\right\}_{k=0}^N
\end{equation}
are the~$2^N$ eigenvectors of~$\mean{\rho_\lambda^{\otimes N}}$ ($\rm perms\{\;\cdot\;\}$ stands for the set of distinct permutations of the set~$\{\;\cdot\;\}$). Defining
\be
\label{ebc01.10.09-3}
A_k\equiv{N \choose k} \int_0^1 d\lambda \,(\epsilon\lambda)^k (1-\epsilon\lambda)^{N-k},
\ee
the eigenvalues of $\mean{\rho_\lambda^{\otimes N}}$ are~ $\nu_k=A_k/{N \choose k}$, and have multiplicity~${N\choose k}$. Therefore,
\be\label{sum2}
\Delta^{\rm comm}=\frac{1}{12}-\frac{1}{4}\sum_k\frac{B_k^2}{A_k}.
\ee
As shown in Appendix~\ref{sums2}, the terms of the sum above
can be written as ratios of Regularized Incomplete Beta Functions thus providing a more
compact expression for the error.
However, we can only give a closed form for~$\Delta$ in the asymptotic limit of very large number of copies.
This requires evaluating the sum in~(\ref{sum2}) up to order $1/N$:
\be\label{sum}
S\equiv\sum_k\frac{B_k^2}{A_k}={1\over3}+{1\over N}\left({4\over3}-{2\over\epsilon}\right)+o(N^{-1})
\ee
(details of this evaluation are also given in Appendix~\ref{form}).
Plugging this expression into~(\ref{sum2}) we obtain
\be
\Delta^{\rm comm}=\frac{1}{N}\left( \frac{1}{2\epsilon} -\frac{1}{3}\right)
+o(N^{-1}).
\label{dcq}
\ee

With the asymptotic techniques introduced in Sec.~\ref{CR} the previous evaluation can be simplified a great deal. Moreover, these techniques enable us to give closed-form expressions of~$\Delta$ for rather more general two-component mixtures.
As already mentioned, the attainability of the CR bound is guaranteed for these (one-parameter) mixtures and its application is particularly simple. From our discussion in~Sec.~\ref{CR}, Eq.~(\ref{ebc28.07.09-1}), we can write
\be
\Delta=\frac{1}{N}\int d\lambda \,\pi(\lambda) \,H_1^{-1}(\lambda)+o(1/N),
\ee
where we recall that~$H_1$ is the QFI of the 1-copy model~(\ref{problem}). As it can be simply read off from~(\ref{ebc03.10.09-1}),
\be
H_1(\lambda)=2\sum_{nm}\frac{|\langle\phi_m|\rho_1-\rho_2|\phi_n\rangle|^2}{\nu_m+\nu_n}.
\ee
Note, however, that $\{\ket{\phi_{n}}\}$ ($\nu_{n}$) are now the eigenvectors (eigenvalues) of $\rho_\la$, rather than of~$\langle\rho_\la\rangle$, and the QFI is thus a function of~$\la$. In the Bloch representation we can write
\be
\label{ebc04.10.09-1}
\rho_r=\frac{1}{2}\left(\mathbb{I}+\vec{r_r}\cdot \vec{\sigma}\right) ,\quad r=1,2 ,
\ee
which holds when~$\rho_1$ and~$\rho_2$ are both qubit states, but also when they are pure states in arbitrary dimensions. Since in these cases the two density matrices can be taken to be real, it suffices to consider~$\vec{\sigma}=(\sigma_x,\sigma_z)$. The eigenvalues and eigenvectors of~$\rho_\lambda$ can be written as
\be
|\phi_\pm\rangle\langle\phi_\pm|=\frac{1}{2}\left(\mathbb{I}\pm \frac{\vec{r_\lambda}\cdot \vec{\sigma}}{r_\lambda}\right),\quad \nu_\pm=\frac{1\pm r_\lambda}{2},
\ee
where, as in previous examples, $\vec{r_\lambda}=\lambda \vec{r_1}+(1-\lambda)\vec{r_2}$ is the Bloch vector of $\rho_\lambda$, and we have defined $r_\la=|\vec r_\la|$. After some algebra one finds
\be
\label{ebc02.10.09-1}
H_1(\lambda)=|\vec{r_1}-\vec{r_2}|^2+\frac{[(\vec{r_1}-\vec{r_2})\cdot\vec{r_\lambda}]^2}{1-r_\lambda^2}.
\ee
For pure states,~$\rho_1=|\varphi_1\rangle\langle\varphi_1|$ and~$\rho_2=|\varphi_2\rangle\langle\varphi_2|$ (i.e.,~$r_1=r_2=1$) one can further simplify this expression and write
\be\label{qfipurequbits}
H_1^{\rm pure}(\lambda)=\frac{1-|\langle\varphi_1|\varphi_2\rangle|^2}{\lambda(1-\lambda)} .
\ee
If the prior is assumed to be flat,  $\pi_{\rm flat}(\lambda)=1$, a trivial integration leads to
\be
\Delta^{\rm pure}= \frac{1}{6N(1-|\langle\varphi_1|\varphi_2\rangle|^2)}+o(1/N).
\ee

If $\rho_1$ and $\rho_2$ are not pure, the Bloch representation~(\ref{ebc04.10.09-1}) holds only for qubit states. Assuming the flat prior,  after a lengthy calculation one finds (to leading order in $1/N$)
\begin{align}
&\Delta^{\rm qubit}=\frac{1}{6N}\frac{6-|\vec{r_1}+\vec{r_2}|^2-r_1^2-r_2^2}{|\vec{r_1}-\vec{r_2}|^2-r_1^2r_2^2+(\vec{r_1}\cdot\vec{r_2})^2}\nonumber\\
&=\frac{1}{6N}\frac{3-\tr\rho_1^2-\tr\rho_2^2-\tr\rho_1\rho_2}{\tr\rho_1^2\!+\!\tr\rho_2^2\!-\!\tr\rho_1^2\,\tr\rho_2^2\!-\!(2\!-\!\tr\rho_1\rho_2)\tr\rho_1\rho_2}.\label{qubits}
\end{align}
Recall that for the cases at hand there exist adaptive measurement that attain the above bound. The reader is referred to Appendix~\ref{two-step} for a specific illustration of this general result.

Before ending this section, we come back to the two commuting states example in~Eq.~(\ref{ebc04.10.09-2}), for which the estimation error, Eq.~(\ref{dcq}), was worked out entirely in the Bayesian framework and the limit $N$ was taken afterwards. The same estimation error can be obtained applying the pointwise CR~result~(\ref{qubits}). It is straightforward to check that this much less costly procedure leads to the same result~(\ref{dcq}), as it should. Recall, however, that it leads to sensible results only if the number~$N$ of copies is exceedingly large, whereas the Bayesian approach works for any~$N$.

\section{Conclusions}\label{sec5}

Quantum ensembles embody what in classical statistics is known as finite mixtures, and can thus be viewed as their quantum counterpart. More precisely, we have a quantum finite mixture whenever a signal can be characterized by a density matrix that is the average of a set of {\em known} states (pure or mixed), as is often the case in quantum communication.
In these situations, one wishes to find the probability law that best describes the signal, or in  other words, the weights that define the quantum ensemble. This has been the subject of the present paper, where we have relied on quantum estimation theory, but also broadened the field by proposing new applications and tools.

The  topics addressed in this paper include: the precise definition of quantum finite mixtures, as an extension of finite mixtures to the quantum domain; optimal estimation (of their weights) when a given number of copies of the average state is available for measurement; optimal estimation in the asymptotic regime of large number of copies; and characterization of the (un)identifiability
of quantum mixtures. For each of these topics we have answered the relevant questions and provided useful results, of which we also give some examples of application.

Going into more detail, we have approached
optimality from both the Bayesian and the `pointwise' points of view. In the former, one minimizes an averaged cost function, which we have chosen to be the covariance-type error matrix of the estimation, over a joint probability involving the measurement outcomes as well as the prior knowledge of the weights.
Our key result is~$\boldsymbol\Delta=\boldsymbol\Lambda-\boldsymbol F$ [see~Eq.~\eqref{eq:DeltaFish}]. It states that the error matrix is the intrinsic uncertainty of the weights minus the Fisher information matrix, which quantifies the information gained in the measurement process. This {\em exact} relation, valid for any number of copies, is linear in the Fisher information matrix, in contrast to the Cram\'er-Rao bound, where the error is lower bounded by the inverse of the Fisher information matrix.
From our relation one obtains a measurement independent lower bound on the error matrix in terms of the Quantum Fisher Information. In those cases where the Braunstein-Caves inequality (which states that the~Fisher Information matrix is upper bounded by the Quantum Fisher Information) is saturated our bound is attainable {\em for any number of copies}. When this holds (e.g., two-component mixtures), we give the optimal measurement protocol, which turns out to be of von Neumann type.

As to the pointwise approach to quantum mixture estimation, we have briefly introduced the Quantum Cram\'{e}r-Rao and the Holevo bounds in the specific context  at hand. We have next applied these tools to obtain lower bounds for the error matrix
of the weights when the number of copies of the average state is asymptotically large.
In those situations, the Bayesian approach becomes rather involved and it is advisable to switch to the tools under discussion. Although the Quantum Cram\'er-Rao and the Holevo bounds can be applied to unidentifiable mixtures, its use requires some technicalities that we have commented upon and illustrated with an example. As one would expect, the accuracy of the weight estimation for such mixtures does not vanish even if an infinite number of copies were available.  A discussion on the relationship between the Bayesian and pointwise approaches has been also given, as well as an example illustrating that the two approaches give consistent results.

Among the examples one can find in this paper, we would like to highlight that of a mixture of a number of orthogonal states, which is relevant in the context of channel estimation. For this problem, and assuming a flat prior distribution of weights we have been able to write the minimal square error in a closed form, valid for any number of orthogonal states and any number of copies of the average state.

This paper is mostly devoted to the formalism and general results concerning quantum finite mixtures and the estimation of their weights.
 The examples are chosen for the sake of illustration, rather than for their practical relevance.
As mentioned in the introduction, real applications of our work are, e.g., the characterization of signals in relevant quantum communication problems and the estimation of probabilities with which various errors occur in a given channel.
We have shown that in some instances the bounds we give are attainable by local two-step adaptive measurements. It remains an open question to establish whether or not collective measurements are necessary in the general case.
Future extensions of our work also include the estimation of mixtures of continuous variable systems.

\section*{ACKNOWLEDGEMENTS}

We acknowledge financial support from:
the Spanish MICINN, through the Ram\'on y Cajal program~(JC), contract FIS2008-01236, and project QOIT
(CONSOLIDER2006-00019); from the Generalitat de
Catalunya CIRIT, contract  2009-SGR-985; and from Alianza 4 Universidades program(JIdV).

\appendix

\section{\boldmath The $\sigma_\lambda$ are physical states}\label{effectivestate}

It is clear from its definition in~(\ref{effsta}) that $\tr\, \sigma_\Blmda=1$. So~$\sigma_\lambda$  is a proper density matrix if $\sigma_\Blmda\geq0$. To prove this inequality, we take any  state $|\psi\rangle$ and define  $p_\Blmda^\psi=\langle\psi|\rho_\Blmda|\psi\rangle\geq0$.  Recalling Eqs.~(\ref{eq:rhog}) and~(\ref{effsta}), we see that the relation~$\langle\psi|\sigma_\Blmda|\psi\rangle\geq0$ is equivalent to
\be
\langle p_\Blmda^\psi\rangle\left[1+\sum_r\left(\lambda_r-\langle\lambda_r\rangle\right)
\left(\frac{\langle \lambda_r p_\Blmda^\psi\rangle}{\langle p_\Blmda^\psi\rangle}-\langle\lambda_r\rangle\right)\right]\ge0.
\label{ebc22.12.09-2}
\ee
[Note that $\langle \lambda_r p_\Blmda^\psi\rangle/\langle p_\Blmda^\psi\rangle\ge0$ and $\sum_r\langle \lambda_r p_\Blmda^\psi\rangle/\langle p_\Blmda^\psi\rangle=1$.]
But~(\ref{ebc22.12.09-2}) immediately follows from the inequality
\be
({\boldsymbol x}-{\boldsymbol z})\cdot({\boldsymbol y}-{\boldsymbol z})\ge -1/2 ,
\label{ebc22.12.09-3}
\ee
where ${\boldsymbol x}=\{x_r\}$, ${\boldsymbol y}=\{y_r\}$ and ${\boldsymbol z}=\{z_r\}$ stand for any three probability vectors.

For the sake of completeness, we also prove~(\ref{ebc22.12.09-3}).
We just need to notice that $({\boldsymbol x}-{\boldsymbol z}) \cdot ({\boldsymbol y}-{\boldsymbol z})$ as a function of~${\boldsymbol z}$ has a minimum at ${\boldsymbol z}_0=({\boldsymbol x}+{\boldsymbol y})/2$. %Since $z_0$ is also a probability vector, we can write $\forall z$
For any ${\boldsymbol x}$, ${\boldsymbol y}$ and~${\boldsymbol z}$ we can thus write
\begin{eqnarray}
({\boldsymbol x}-{\boldsymbol z})\cdot({\boldsymbol y}-{\boldsymbol z})&\geq&({\boldsymbol x}-{\boldsymbol z}_0)\cdot({\boldsymbol y}-{\boldsymbol z}_0)\nonumber\\
&=&- \frac{|{\boldsymbol x}-{\boldsymbol y}|^2}{4}\geq-\frac{1}{2},
\end{eqnarray}
which is the inequality~(\ref{ebc22.12.09-3}).

\section{Useful formulae}\label{form}

\subsection{Averages with a flat prior}\label{aver}

Recall our notation:  $d\Blmda=\delta\left(\sum_r\lambda_r-1\right)\prod_r d\lambda_r$.
Then, one can prove the following useful result
\be
\int d \Blmda\;
\prod_r \lambda_r^{k_r}
={k_1!\cdots k_M!\over \left(M-1+\sum_r k_r\right)!},
\label{ebc18.07.09-4}
\ee
where the integration is restricted to positive values of~$\lambda_r$, $r=1,\dots, M$.
Although we use this integral for $k_r$ being positive integers, the result can be generalized to complex~$k_r$ by simply replacing the factorials by Euler Gamma functions: $k!\to\Gamma(k+1)$.

In particular,  Eq.~(\ref{ebc18.07.09-4}) and the normalization condition $\int d\Blmda \pi(\Blmda)=1$ imply that the flat distribution is given by
\be
\pi_{\rm flat}(\Blmda)=(M-1)!\,.
\ee

\subsection{Sums
}\label{sums}

Some results in~Sec.~\ref{febc18.07.09-1} require computing sums of the form
$\sum_{\boldsymbol{k}, r} f(k_r)$, where $k_r$ are the components of the vector~$\boldsymbol k$. They are positive integers that add up to $N$, and the sum extends over all the
\be
{M+N-1\choose M-1}
\label{ebc18.07.09-5}
\ee
such vectors. To compute this sums, we first note that
\be
\sum_{\boldsymbol{k}, r} f(k_r)=\sum_{r=1}^M\left(\sum_{\boldsymbol{k}} f(k_r)\right)=M \sum_{\boldsymbol{k}} f(k_1),
\label{ebc18.07.09-6}
\ee
where we have used that the sum in parenthesis is independent of~$r$. This is so because the set of all vectors~$\boldsymbol k$ is invariant under  $k_r\to k_{\sigma(r)}$, where $\sigma$ is any permutation of the symmetric group~$S_M$, and thus $\sum_{\boldsymbol{k}} f(k_r)=\sum_{\boldsymbol{k}} f(k_{\sigma(r)})$.
We next note that any vector~$\boldsymbol k$ whose first component is fixed to be~$k_1$ gives the same contribution, $f(k_1)$, to the last sum in~(\ref{ebc18.07.09-6}).
The number of such vectors follows from~(\ref{ebc18.07.09-5}) by simply making the substitutions~$M\to M-1$ and~$N\to N-k_1$.
Hence,
\be
\sum_{\boldsymbol{k}, r} f(k_r)=M\sum_{k_1=0}^N {M+N-k_1-2\choose M-2} f(k_1).
\ee

For the particular case we need in~Sec.~\ref{febc18.07.09-1}, $f(x)=x^2$ and the corresponding sum gives
\be
\sum_{\boldsymbol{k},r}k_r^2={2N+M-1\over M+1}{(M+N-1)!\over(M-1)!(N-1)!} .
\ee

\subsection{Evaluation of the sum (\ref{sum})}\label{sums2}

Recalling the definitions~(\ref{ebc01.10.09-2}) and~(\ref{ebc01.10.09-3}), and after some algebra,
we have
\begin{eqnarray}
S&=&{1\over N+1}{4\over\epsilon^3}\sum_{k=0}^N\left({k+1\over N+2}\right)^2{I^2_\epsilon(k+2,\bar k+1)\over I_\epsilon(k+1,\bar k+1)} -1\nonumber
\\
&\equiv&{R(\epsilon)\over \epsilon^3}-1 ,
\label{ebc01.10.09-4}
\end{eqnarray}
where $\bar k\equiv N-k$ and  $I_x(a,b)$ is the Regularized Incomplete Beta Function,
\be
\label{RIBF}
I_x(a,b)={B_x(a,b)\over B(a,b)}={1\over B(a,b)}\int_0^x \!\! dt\, t^{a-1}(1-t)^{b-1} .
\ee
To obtain~(\ref{ebc01.10.09-4}) we have also  used that
\be
\sum_{k=0}^N I_\epsilon(k+1,\bar k+1)=(N+1)\epsilon
\ee
and
\be
\sum_{k=0}^N {k+1\over N+2} I_\epsilon(k+1,\bar k+1)=(N+1){\epsilon^2\over2} ,
\ee
which both follow immediately from the definition in Eq.~(\ref{RIBF}). Recall also that $B_1(a,b)\equiv B(a,b)$, where $B(a,b)$ is the standard (complete) Beta Function, $B(a,b)=\Gamma(a)\Gamma(b)/\Gamma(a+b)$.
According to the Euler-MacLaurin formula, the sum in~(\ref{ebc01.10.09-4}) can be approximated by an integral which, after differentiating with respect to $\epsilon$, can be cast as
\begin{eqnarray}
R'(\epsilon)&=&{4N\over N+1}\int_0^1 dx\,\left({N x+1\over N+2}\right)^2 { I_\epsilon(Nx+2,N\bar x+1)\over I_\epsilon(Nx+1,N\bar x+1)}\nonumber\\
&\times&
\left\{
{2(N+2)\epsilon\over Nx+1}-
{ I_\epsilon(Nx+2,N\bar x+1)\over I_\epsilon(Nx+1,N\bar x+1)}
\right\}\nonumber\\
&\times&
{\epsilon^{N x}(1-\epsilon)^{N\bar x}\over B(N x+1,N\bar x+1)} ,\label{rprima}
\end{eqnarray}
where $\bar x\equiv 1-x$. The last factor peaks at  $x=\epsilon$ as $N$ becomes large and can be replaced by the Gaussian
$$
{N+1\over N} \sqrt{{N\over 2\pi \epsilon(1-\epsilon)}}\exp\left\{-N{(x-\epsilon)^2\over 2\epsilon(1-\epsilon)}\right\}  .
$$
Since we are interested only in terms that vanish asymptotically as~$N^{-1}$,
we can drop those that vanish exponentially, and approximate~Eq.\ (\ref{rprima}) by
\begin{eqnarray}
R'(\epsilon)\!\!&=&\!\!4\!\int_{-\infty}^\infty \kern-.5em dx\left({N x+1\over N+2}\right)^2
\!\!\left\{\!
{2(N+2)\epsilon\over Nx+1}-
1
\!
\right\}\nonumber\\
\!\!&\times&\!\!
\sqrt{{N\over 2\pi \epsilon(1-\epsilon)}}\exp\left\{-N{(x-\epsilon)^2\over 2\epsilon(1-\epsilon)}\right\} .
\label{ebc01.10.09-6}
\end{eqnarray}
For the same reason, we can expand the first line in~(\ref{ebc01.10.09-6}) up to first order in~$u^2\equiv(x-\epsilon)^2$ and write
\begin{eqnarray}
R'(\epsilon)\!\!&=&\!\!4\int_{-\infty}^\infty du\
(\epsilon^2-u^2)\nonumber\\
\!\!&\times&\!\!
\sqrt{{N\over 2\pi \epsilon(1-\epsilon)}}\exp\left\{\!-N{u^2\over 2\epsilon(1-\epsilon)}\!\right\} .
\end{eqnarray}
The remaining integral gives
\be
R'(\epsilon)=4\epsilon^2-{4\over N}\epsilon(1-\epsilon)  .
\ee
Hence,
\begin{eqnarray}
R(\epsilon)&=&R(0)+\int_0^\epsilon ds\, R'(s)\nonumber\\
&=&{4\over3}\epsilon^3-{2\over 3N}\epsilon^2(3-2\epsilon) ,
\end{eqnarray}
from which the final result follows.

\section{Two-step adaptive measurement in the asymptotic limit}\label{two-step}

In this appendix we give an explicit example of the two-step adaptive measurement protocol  that attains the Cram\'er-Rao bound asymptotically [see~Sec.~\ref{CR}, the paragraph after Eq.~(\ref{ebc28.07.09-3})]. To ease the calculation we choose  the simplest instance: that of a mixture of two pure states,~$\rho_\lambda=\lambda\rho_1+(1-\lambda)\rho_2$.
This mixture has been already considered
in Sec.~\ref{sec4}, in the paragraph after~Eq.~(\ref{ebc02.10.09-1}). Here we stick to the same notation.
If~$\rho_{r}$~($r=1,2$) are pure, without loss of generality they can be chosen to be
\begin{equation}
\rho_r={1\over2}\left[\openone+\sigma_z\cos\theta+(-1)^{r+1}\sigma_x\sin\theta\right)]
\end{equation}
(as if they were qubit states on the equator  of the Bloch sphere),
where $\cos\theta=\sqrt{\tr\,\rho_1\rho_2}=|\langle\varphi_1|\varphi_2\rangle|$ is the overlap.

Let us assume that we are given~$N$ copies of the state~$\rho_\lambda$. On a first stage of the protocol, we take~$\sqrt{N}$ of these copies and perform on each of them  a same measurement, with the aim of obtaining an initial, rough estimate of $\lambda$, which we denote by~$\lambda_{\rm ini}$. Since these measurements use uncorrelated copies and are themselves independent, we expect to benefit from the well understood statistical  improvement that results from averaging over the $\sqrt N$ samples. Thus, we can assume that, in average,
\begin{equation}\label{stat_impr}
(\lambda-\lambda_{\rm ini})^2\sim \alpha/\sqrt N,
\end{equation}
where $\alpha$ is some constant whose value depends on the precise measurement that we perform.

On a second stage, we refine the rough estimation obtained in the preceding stage by performing a (nearly optimal) measurement on the remaining $N-\sqrt{N}$ copies. As~discussed in~Sec.~\ref{CR} (See also~Sec.~\ref{ebc28.07.9S-1}), the optimal measurement is described by the set of projector,~$\{P_\chi(\lambda)\}$ (it is a von~Neumann measurement), onto the different eigenspaces of the~SLD, $L(\lambda)$, of our model evaluated at~$\lambda$.
For our example, one can readily find that
\be
\label{SLD_ex}
L(\lambda)={(1-2\lambda)(\openone+\sigma_z\cos\theta)+\sigma_x\sin\theta \over2\lambda(1-\lambda)}  .
\ee
However, since we do not know the true value of~$\lambda$, we choose the measurement to be given by~$\{P_\chi(\lambda_{\rm ini})\}$, and hope this change will not affect optimality. Let us check that this is indeed the case. To this end, we diagonalize~(\ref{SLD_ex}), obtain $\{P_\chi(\lambda_{\rm ini})\}$ and, in turn, compute its Fisher information defined in~(\ref{eq:fisher}). We obtain
\begin{equation}
F_1={\sin^2\theta\over \la(1-\la)+(\la_{\rm ini}-\la)^2\cos^2\theta} .
\end{equation}
(recall that the subscript~$1$ refers to one copy).
Thus, the error  of performing this measurement on the $N-\sqrt N$ copies  is
\be
\Delta(\lambda)\!=\!{1\over N-\sqrt N}\left\{{\la(1-\la)\over \sin^2\theta}+(\la_{\rm ini}\!-\!\la)^2\cot^2\theta \right\}.
\ee
For sufficiently large $N$ (so that $\sqrt N$ itself is also very large), Eq.~(\ref{stat_impr}) holds in average, and
\begin{eqnarray}
\Delta(\lambda)\!\!&=&\!\!{\la(1-\la)\csc^2\theta\over N(1-N^{-1/2}) }+{\alpha \cot^2\theta \over N^{3/2}(1-N^{-1/2}) }\nonumber\\[.5em]
\!\!&=&\!\!
{\la(1-\la)\csc^2\theta\over N}+O(N^{-3/2}) ,
\end{eqnarray}
thus attaining the optimal bound, as can be read off from~(\ref{qfipurequbits}).

\end{document}